\documentclass[%
preprint,
showpacs,preprintnumbers,
 amsmath,amssymb,
 aps,
]{revtex4-1}

\usepackage{graphicx}% Include figure files
\usepackage{dcolumn}% Align table columns on decimal point
\usepackage{bm}% bold math
\begin{document}

%\preprint{APS}

\title{Localization of low-frequency oscillations in single-walled carbon nanotubes. \\
Supplementary materials}% Force line breaks with \\

\author{V.V. Smirnov}%
\email{vvs@polymer.chph.ras.ru}
\author{D.S. Shepelev}
\author{L.I. Manevitch}
 
\affiliation{%
 Institute of Chemical Physics, RAS, Moscow, Russia\\
119991, 4 Kosygin str., Moscow, Russia
}%

\date{\today}% It is always \today, today,
\maketitle

%\begin{document}

In this supplement we present a short derivation of dynamical equations of  thin elastic shell  in the framework of modified nonlinear Sanders-Koiter theory, and their relationship with the dynamics of one-dimensional oscillatory chain.

The dynamics of carbon nanotubes (CNT) is one of the few areas of solid state physics, in which the classical theory of thin elastic shells (TTES) can be legitimately applied. It is noteworthy that, in contrast to macroscopic mechanics, where the fundamental limits of TTES are restricted by the possibility of plastic deformation, this theory can also be used for large displacements of CNTs, even in the analysis of its collapse.
The only complicating factor is the uncertainty of the parameter characterizing the thickness of the CNT.
%, which determines along with elastic moduli, its macroscopic stiffness. Therefore, that should be determined directly by atomic interaction potentials. In addition, for the investigation of the CNT collapse it is necessary to consider the energy manifested in coming together of opposite CNT regions.
The applicability of a well-designed TTES allows us to obtain an effective description of the vibrational spectrum in the framework of the linear approximation. 
 It is essential, however, that a theoretical analysis is available only for the simplest  of boundary conditions, when the CNT of finite size can be considered as part of an infinite CNT. However, the modified theory presented below admits efficient study of nonlinear dynamics of CNT.

\section{Shell theory equations}

We use the dimensionless variables which determine the elastic deformation of circular thin shell. All displacements ($u$ - longitudinal along the CNT axis, $v$ - tangential and $w$ - radial displacement, respectively) are measured in the units of CNT radius $R$. The coordinate along the CNT axis  $\xi=x/L$ is measured via the length of nanotube and variates from $0$ up to $1$, and $\theta$ is the azimuthal angle. The displacements and respective deformations refer to the middle surface of the shell. One can define the dimensionless energy and time  variables, which are measured in the units $E_{0}=YRLh/(1-\nu^{2})$ and $t_{0}=1/\sqrt{Y/\rho R^{2}(1-\nu^{2})}$, respectively. Here $Y$ is the Young modulus of graphene sheet, $\rho$ - its mass density, $\nu$ - the Poisson ratio of CNT, and $h$ is the effective thickness of CNT wall. There are two dimensionless geometric parameters which characterize CNT: the first one is inverse aspect ratio $\alpha=R/L$ and the second - effective thickness shell $\beta=h/R$.

The energy of elastic deformation of CNT in the dimensionless units is written as follows:

\begin{equation}
\begin{split}
E_{el} =\frac{1}{2} \int\limits_{0}^{1} \int\limits_{0}^{2 \pi}\left(\varepsilon_{\xi}^{2} + \varepsilon_{\theta}^{2} + 2 \nu \varepsilon_{\xi} \varepsilon_{\theta} + \frac{1 - \nu}{2} \varepsilon_{\xi \theta}^{2}\right)d\xi d\theta +\\
+\frac{\beta^{2}}{24} \int\limits_{0}^{1} \int\limits_{0}^{2 \pi}\left(\kappa_{\xi}^{2} + \kappa_{\theta}^{2} + 2 \nu \kappa_{\xi} \kappa_{\theta} + \frac{1 - \nu}{2} \kappa_{\xi \theta}^{2}\right)d\xi d\theta ,
\label{energy_elst}
\end{split}
\end{equation}

where $\varepsilon_{\xi}$, $\varepsilon_{\theta}$ and $\varepsilon_{\xi \theta}$ are the longitudinal, circumferential and shear deformations, and $\kappa_{\xi}$, $\kappa_{\theta}$ and $\kappa_{\xi \theta}$ are the longitudinal and circumferential curvatures, and torsion, respectively. One should note that both curvatures and torsion are the dimensionless variables in accordance with our definition of dispacement field $(u, v, w)$.

The dimensionless kinetic energy is equal to

\begin{equation}
\begin{split}
E_{kin} =\frac{1}{2} \int\limits_{0}^{1} \int\limits_{0}^{2 \pi}\left(\dot{u}^{2} + \dot{v}^{2} + \dot{w}^{2}\right)d\xi d\theta  ,
\label{energy_kin}
\end{split}
\end{equation}

where the point over the symbol corresponds to the partial derivative with respect to dimensionless time $t$.

Using the Sanders-Koiter approximation for the shell deformations and curvatures of defectless thin shell  we can write

\begin{equation}
\begin{split}
\varepsilon_{\xi} = \alpha \frac{\partial u}{\partial \xi} + \frac{\alpha^{2}}{2}( \frac{\partial w}{\partial \xi})^{2} +\frac{1}{8}(\alpha \frac{\partial v}{\partial \xi}-\frac{\partial u}{\partial \theta})^{2} \\ 
\varepsilon_{\theta} = \frac{\partial v}{\partial \theta} + w + \frac{1}{2} (\frac{\partial w}{\partial \theta} -v)^{2}+\frac{1}{8}(\frac{\partial u}{\partial \theta}-\alpha \frac{\partial v}{\partial \xi})^{2}  \\
 \varepsilon_{\xi \theta} = \frac{\partial u}{\partial \theta} + \alpha \frac{\partial v}{\partial \xi} +\alpha \frac{\partial w}{\partial \xi}(\frac{\partial w}{\partial \theta}-v) 
\end{split}
\label{deformation}
\end{equation}
\begin{equation}
\kappa_{\xi} = - \alpha^{2} \beta \frac{\partial^{2} w}{\partial \xi^{2}},\ \ \kappa_{\theta} = \beta \left(\frac{\partial v}{\partial \theta} - \frac{\partial^{2} w}{\partial \theta^{2}} \right),\ \ 
\kappa_{\xi \theta} = \beta \left( - 2 \alpha \frac{\partial^{2} w}{\partial \xi \partial \theta} + \frac{3 \alpha}{2} \frac{\partial v}{\partial \xi} - \frac{1}{2} \frac{\partial u}{\partial \theta}\right).
\label{curvation}
\end{equation}

Substituting the eqs.(\ref{deformation}, \ref{curvation}) into eq.(\ref{energy_elst}) one can express the Lagrange function

\begin{equation}
 \it{L}=E_{kin}-E_{el} 
\label{eq:Lgr}
\end{equation}

via the displacement field $(u, v,  w)$. The variation of Lagrange function (\ref{eq:Lgr}) with respect to the displacements leads to the system of equations, the linear approximation of which determines a full vibration spectrum of CNT. Unfortunately, these equations are too difficult to solve  analytically even in the framework of linear approximation.  But if we can make some physically founded relationships between the displacement components one can try to simplify the CNT dynamics description. We consider the low-frequency optical-type vibrations which are specified by circumferential wave number $n=2$:

\begin{equation}
\begin{split}
u(\xi, \theta, t)=U_{0}(\xi,t)+U(\xi,t) \cos( 2\theta) \\
v(\xi, \theta, t)=V(\xi,t) \sin(2\theta) \\
w(\xi, \theta, t)=W_{0}(\xi,t)+W(\xi,t) \cos(2 \theta)
\end{split}
\label{eq:var1}
\end{equation}

 This branch is characterized by relatively small circumferentional and shear deformations, while the displacements themselves may not be small. In such a case we can write:

 $$\varepsilon_{\theta}=0 ; \varepsilon_{\xi \theta}=0 $$

These relations allow us to express the longitudinal and tangential components, and axially symmetric part of the radial displacement via the radial one. These relationships can be written as folows:

\begin{equation}
\begin{split}
V(\xi,t)=-\frac{1}{2} W(\xi,t);   
\frac{\partial U(\xi,t)}{\partial \xi}=-\frac{\alpha}{4} W(\xi, t) \\
W_{0}(\xi,t)=-\frac{1}{16}(9W^{2}(\xi,t)+\alpha^{2} (\frac{\partial W(\xi,t)}{\partial \xi})^{2});    
\frac{\partial U_{0}}{\partial \xi}=-\frac{5}{16} \alpha (\frac{\partial W(\xi,t)}{\partial \xi})^{2}
\end{split}
\label{eq:uv2w}
\end{equation}

Omitting the calculation details one can write the final equation of motion in the terms of radial displacement $W(\xi, t)$:

\begin{equation}
\begin{split}
 \frac {\partial^{2}W}{\partial t^{2}}+ \frac{\beta^{2}}{12} \frac{n^{2} (n^{2}-1)^{2}}{n^{2}+1} W -\frac{\alpha^{2} \beta^{2} (n^{2}-1)(n^{2}-1+\nu)}{6 (n^{2}+1)} \frac{\partial ^{2} W}{\partial \xi^{2}} - \frac{\alpha ^{2}}{n^{4}} \frac{\partial ^{4} W}{\partial \xi^{2} \partial t^{2}}+\frac{\alpha^{4} (12+n^{4} \beta^{2})}{12 n^{2}(n^{2}+1)} \frac{\partial^{4} W}{\partial \xi^{4}} \\
+\frac{(n^{2}-1)^{4}}{2 n^{2} (n^{2}+1)}  (W (\frac{\partial W}{\partial t} )^{2}+W^{2} \frac{\partial ^{2} W}{\partial t^{2}}) - 
\frac{\alpha ^{2} (n^{2}-1)^{2}}{2 n^{2} (n^{2}+1)}  \biggl[ \frac{\partial ^{2} W}{\partial t^{2}} (\frac{\partial W}{\partial \xi})^{2} +2 \frac{\partial W}{\partial t} \frac{\partial W}{\partial \xi} \frac{\partial ^{2} W}{\partial \xi \partial t} -  \\
W (\frac{\partial ^{2} W}{\partial \xi \partial t})^{2}+(\frac{\partial W}{\partial t})^{2} \frac{\partial ^{2} W}{\partial \xi^{2}} +W \frac{\partial ^{2} W}{\partial t^{2}} \frac{\partial ^{2} W}{\partial \xi^{2}} + 4 \alpha ^{2} (\frac{\partial W}{\partial \xi})^{2} \frac{\partial ^{2} W}{\partial \xi^{2}}\biggr]- \\
\frac{\alpha^{4}}{2 n^{2} (n^{2}+1)} \Biggl[  (\frac{\partial ^{2} W}{\partial \xi \partial t})^{2} \frac{\partial ^{2}W}{\partial \xi^{2}}+2 \frac{\partial W}{\partial \xi}\frac{\partial ^{3} W}{\partial \xi \partial t^{2}} \frac{\partial^{2} W}{\partial \xi^{2}}-2 \alpha^{2} (\frac{\partial^{2} W}{\partial \xi^{2}})^{3}+2 \frac{\partial W}{\partial \xi} \frac{\partial^{2} W}{\partial \xi \partial t} \frac{\partial^{3} W}{\partial \xi^{2} \partial t} + \\ (\frac{\partial W}{\partial \xi})^{2}\frac{\partial^{4} W}{\partial \xi^{2} \partial t^{2}}-8 \alpha^{2} \frac{\partial W}{\partial \xi} \frac{\partial^{2} W}{\partial \xi^{2}} \frac{\partial^{3} W}{\partial \xi^{3}} - \alpha^{2} (\frac{\partial W}{\partial \xi})^{2} \frac{\partial^{4} W}{\partial \xi^4}  \Biggr] = 0,
\end{split}
\label{eq:WWW}
\end{equation}

where $n$ has to be equal to $2$. The linearization of eq.(\ref{eq:WWW}) leads to the spectrum of natural vibrations at the periodic boundary conditions:

\begin{equation}
\omega^{2}=\frac {n^{4} (n^{2} - 1)^{2}\beta^{2} + \alpha^{2} \beta^{2}\pi^{2} n^{2} (n^{2} - 1) (n^{2} - 1 + \nu) k^{2} + \alpha^{4}\pi^{4} (12 + n^{4}\beta^{2}) k^{4}} {12 (n^{4} + n^{2} + \alpha^{2}\pi^{2} k^{2})},
\label{eq:omega}
\end{equation}

where $k$ is a longitudinal wave number corresponding to the number of half-wave along the CNT axis.

\begin{figure}[h!]
\center{\includegraphics[width = 100mm, height = 60mm]{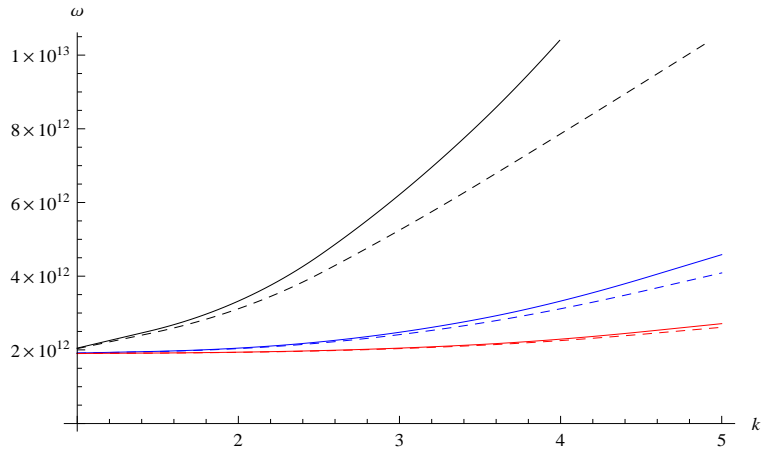}}
\caption{Linear approximation spectra for simply supported CNT calculated: according to eq.(\ref{eq:omega}) (solid curves) and full equations of Sanders-Koiter theory (dashed curves). Aspect ratios: 10 (black), 20 (blue), 30 (red). $\omega$ in $s^{-1}$.}
\label{fig:spectrum1}
\end{figure}

Eq.(\ref{eq:WWW}) allows us to calculate the natural frequencies in the linear approximation as well as to estimate the effect of nonlinearity on these frequencies at different boundary conditions.  Fig.\ref{fig:spectrum1} demonsrates the comparison between linear spectra calculated according to eq.(\ref{eq:omega}) and full equations of Sanders-Koiter theory. One can see their good agreement in the low-frequency region which is of the main interest for analysis of nonlinear low-energy problems.

Assuming that the solution of eq.(\ref{eq:WWW}) under periodic boundary conditions  is represented as follows:

\begin{equation}
 W(\xi,t)=f_{1}(t) sin(\pi \xi)+f_{2}(t) sin(2 \pi \xi) ,
\label{eq:solution}
\end{equation}

we can get the equations of motion for the modal amplidutes $f_{1}$ and $f_{2}$, using the Galerkin procedure. As the result we get two ordinary differential equations for the functions $f_{1}$ and $f_{2}$: 

\begin{equation}
\begin{split}
(20 + \pi^{2} \alpha^{2}) f_ {1}^{\prime\prime} + \frac {1} {3} (36 \beta^{2} + \pi^{4} \alpha^{4} (3 + 4\beta^{2}) + 6\pi^{2}\alpha^{2}\beta^{2} (3 + \nu))f_{1}+\frac{1}{2}\pi^{4}\alpha^{4}(\pi^{2}\alpha^{2}-9)f_{1}^{3}+ \\ 2\pi^{4}\alpha^{4}(5\pi^{2}\alpha^{2}-18)f_{1}f_{2}^{2}+\frac{3}{8}(81+6\pi^{2}\alpha^{2}+\pi^{4}\alpha^{4})(f_{1}f_{1}^{\prime 2}+f_{1}^{2}f_{1}^{\prime \prime})+ \\ \frac{1}{4}(81+\pi^{4}\alpha^{4})(2f_{1}f_{1}^{\prime}f_{2}^{\prime}+f_{2}^{2}f_{1}^{\prime \prime}) + \frac{1}{4} (162+45\pi^{2}\alpha^{2}+8\pi^{4}\alpha^{4})f_{1}f_{2}f_{2}^{\prime \prime}=0 
\end{split}
\label{eq:mode1}
\end{equation}

\begin{equation}
\begin{split}
(5 + \pi^{2} \alpha^{2}) f_ {2}^{\prime\prime} + \frac {1} {3} (9 \beta^{2} + 4\pi^{4} \alpha^{4} (3 + 4\beta^{2}) + 6\pi^{2}\alpha^{2}\beta^{2} (3 + \nu))f_{2}+\frac{1}{2}\pi^{4}\alpha^{4}(5\pi^{2}\alpha^{2}-18)f_{1}^{2}f_{2}+ \\ 
2\pi^{4}\alpha^{4}(4\pi^{2}\alpha^{2}-9)f_{2}^{3}+\frac{1}{16}(81+45\pi^{2}\alpha^{2}+4\pi^{4}\alpha^{4})(f_{2}f_{1}^{\prime 2}+f_{1}f_{2}f_{1}^{\prime \prime})+ \\
\frac{1}{16}(81+4\pi^{4}\alpha^{4})(2f_{1}f_{1}^{\prime}f_{2}^{\prime}+f_{1}^{2}f_{2}^{\prime \prime}+f_{1}f_{2}f_{1}^{\prime \prime}) + 
\frac{3}{32} (81+24\pi^{2}\alpha^{2}+16\pi^{4}\alpha^{4})(f_{2}f_{2}^{\prime 2}+f_{2}^{2}f_{2}^{\prime \prime})=0
\end{split}
\label{eq:mode2}
\end{equation}

Using the multiscale asymptotic expansion one can obtain  eqs. (2) of the Letter.

\section{1D chain}

\begin{figure}%[hbtp]

\noindent
\centering{
 \includegraphics[width=80 mm]{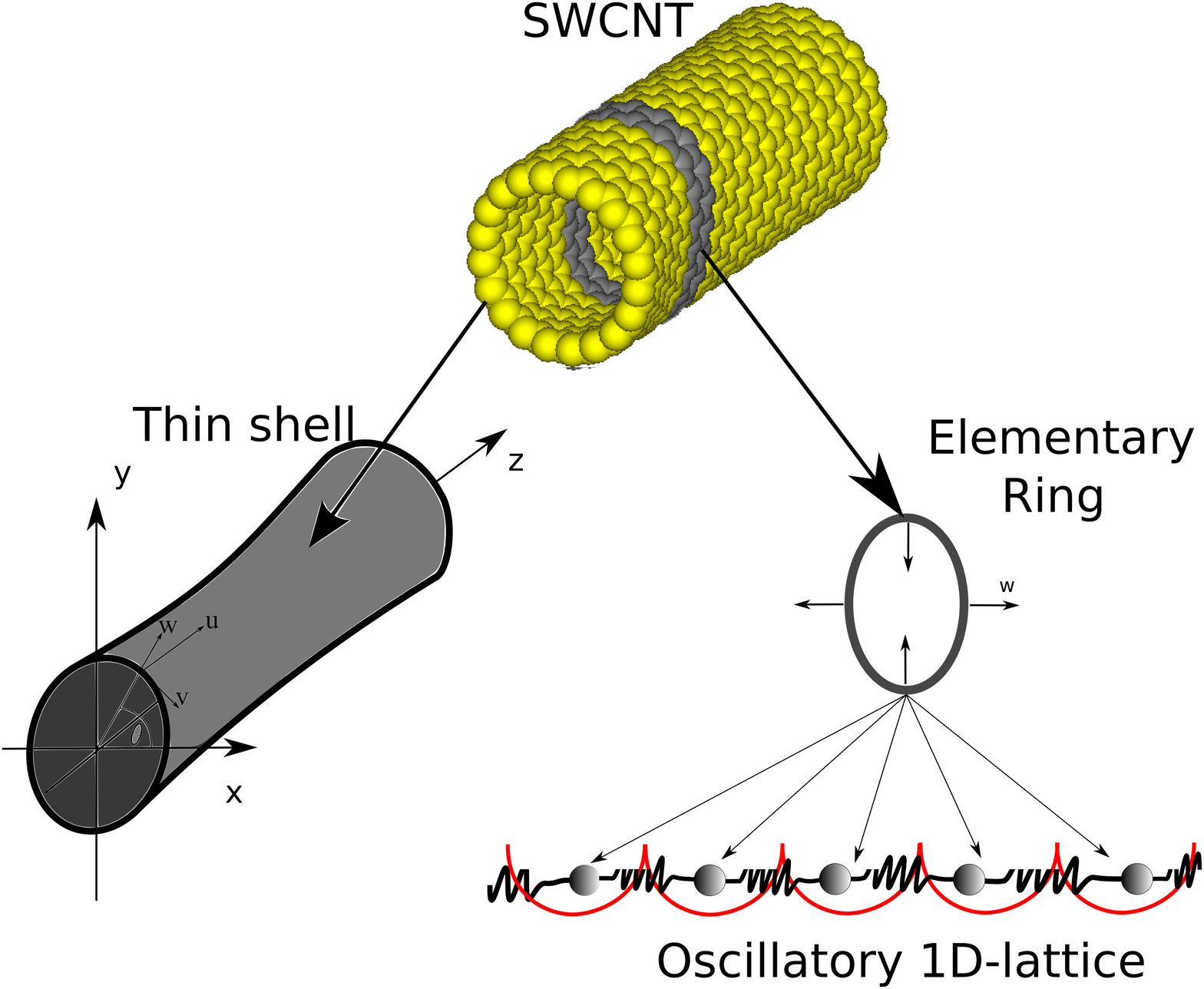}
 \includegraphics[width=80 mm]{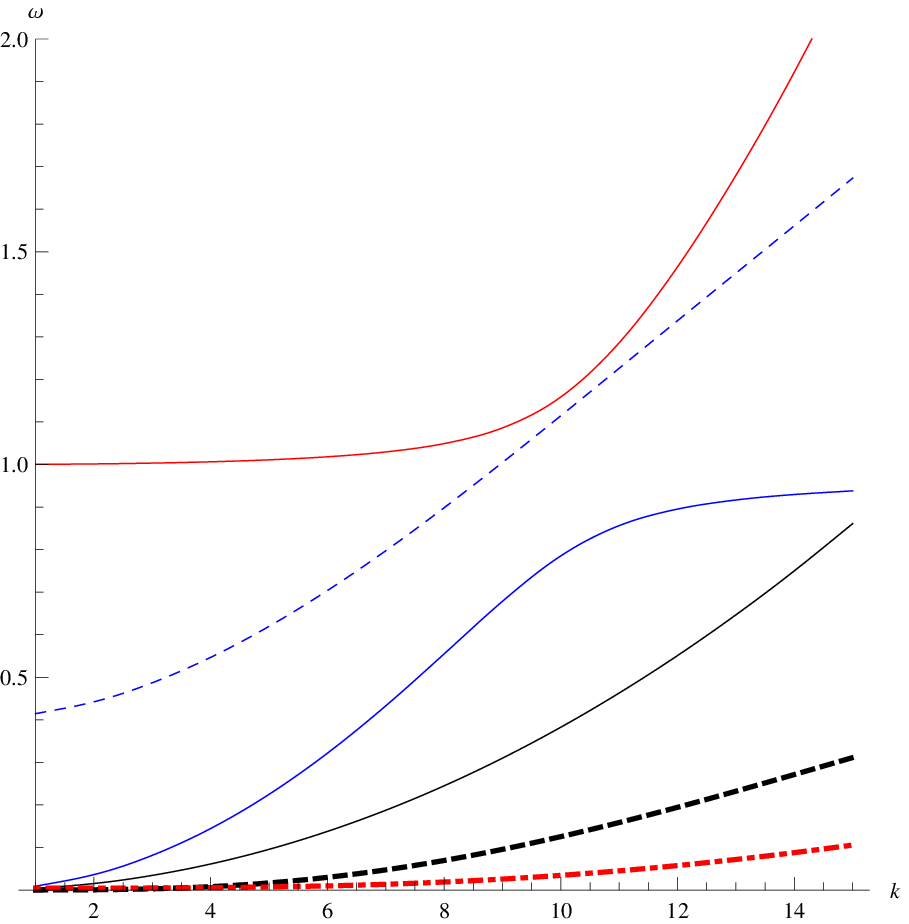}
}
\caption{
The left panel: SWCNT presentations as a thin cylindrical (left part) and a one-dimensional oscillatory chain (right part). In the oscillatory chain the red curves show the substrate potential that results from the  rigidity of an elementary ring itself. The right panel contains the CNT spectrum according to the exact Sanders-Koiter thin shell theory: solid curves correspond to circumferential number $n=0$, dashed ones - to $n=1$ and dot-dashed curve - to $n=2$.}
\label{fig:CNT2chain}

\end{figure}

We consider the localization of low-frequency optical vibrations of CNTs. To elucidate  the origin of this effect in the framework of thin shell theory, one can use a simple qualitatively analogy between CNT and a one-dimensional oscillatory chain (see the figure \ref{fig:CNT2chain}). Let us virtually cut the CNT into "elementary" rings across nanotube axis. Assuming that each ring can be deformed in its plane only, one can consider some simple modes of deformation, which are characterized by the azimuthal wave number $n$. Then each ring itself is similar to some oscillator, the natural frequency of which is determined by the ring rigidity with respect to deformation of its shape. 
The bonds between neighboring rings are provided with the flexural rigidity of CNT's generatrix. Selecting certain mode of the ring deformation, one can construct some equivalent one-dimensional oscillatory chain with the on-site potential determined by the bend stiffness of elementary ring, and inter-oscillatory bonds are realized by the longitudinal flexure stiffness of CNT. There are two optical-type branches in the CNT spectrum (see the right panel of fig. \ref{fig:CNT2chain}): the first one contains the well-known Radial Breathing Mode (RBM), which is associated with azimuthal wave number $n=0$ and corresponds to uniform expansion-compression of the ring. The second branch contains modes which relate to $n=2$, and prevalent deformation corresponds to deviation of the ring shape from initial circular one (Circumferential Flexure Mode - CFM). One should note that CFM is the lowest optical mode in the CNT spectrum. The amplitude of radial displacement of elementary ring can be chosen as main variable both for RBM and for CFM. Then the Hamilton function of equivalent one-dimensional chain can be written as follows:

\begin{equation}
\label{1Dchain}
 H=\sum{_k}\frac{\rho}{2}(\frac{dw{_k}}{dt})^2 +U(w{_k})+\frac{c^2}{2}(w_{k+1}-2w_{k}+w_{k-1})^{2} ,
\end{equation}

where $w_k$ is amplitude of the radial displacement of $k$-th ring, $\rho$ is the mass density of ring, $U(w_{k})$ describes the (nonlinear) ring's stiffness and the parameter $c$ is defined by the bend rigidity of CNT.

It is clearly that system (\ref{1Dchain}) possesses the optical-type spectrum:

\begin{equation}
\label{1Dspectrum}
\Omega^{2}=\omega_{0}^{2}+16 c^{2} \sin^{4}(\frac{k}{2}),
\end{equation}
where $\omega_{0}^{2}=d^{2} U(w)/dw^{2}|w=0$ and $k$ is a wave number.
 Such a spectrum, as it was mentioned in the Letter, exhibits the respective crowding near the low-frequency edge, i.e. for normal modes with the minimal wave number.The lowest (by frequency) mode of spectrum (\ref{1Dspectrum}) corresponds to wave number $k=0$ and describes the uniform vibrations of oscillators. A combination of two low-frequency modes can be considered as a weak modulation of lowest mode. In the phase portrait of the system the attraction regions of these modes are separated by the phase trajectory which is denoted as "the Limiting Phase Trajectory". A motion along this trajectory leads to slow energy exchange between different parts of the system. In the nonlinear case (when the amplitudes of modes become large enough) one of modes losses its stability that leads to creation of new type trajectory - the separatrix, passing through the unstable stationary point and surrounding two stationary points, corresponding to nonlinear normal modes splitted from unstable one. As it has been previously shown, the weak energy localization occurs in such a system with  finite chain length. The source of localization is the coincidence of LPT and separatrix. 

\section{MD simulation}

\begin{figure}
\noindent
\centering{
\includegraphics[width=80mm]{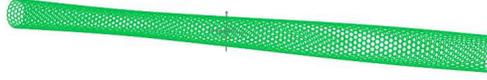}
}
\caption{Snapshot of typical deformation of CNT during the vibration associated with CFM spectrum branch.}
\label{fig:CNT_deform}
\end{figure}

To verify the results of analytical model the simulation of the low-frequency vibrations of carbon nanotubes was performed by  molecular dynamics (MD)  techniques using realistic inter-atomic  potential  functions. 
Classical molecular dynamics technique which uses predefined potential functions (force fields) was applied for the calculation of  the total potential energy of  the system.  
The typical MD experiment was devided into several stages. At the first stage the CNT was kept at  high temperature ($\simeq 400 K$) for structural relaxation. Then the termostat temperature was decreased with a constant rate down to approximately 1 K with a subsequent low-temperature relaxation. The third stage deals with CNT deformation according to analytical solution with subsequent relaxation. The second version of initial conditions consisted in the determination of initial velocities of atoms at zero initial displacements. After that the external field was turned off, and the free natural oscillations of CNT with the fixed boundary conditions were realized. In accordance with analytical description, the atoms at the edges of CNT were fixed by the force field against any radial displacements ($W(0,t)=W(1,t)=0$) that corresponds to the boundary conditions similar to simply supported shell. The typical snapshot of CNT deformation during the MD simulation is shown in fig.(\ref{fig:CNT_deform}). 

The consequent analysis of MD simulation data included  the control of  natural frequencies and energy distribution along the CNT axis via  variation of the oscillation amplitude. The 3D pictures of energy distribution along the CNT axis measured during the MD simulations are shown in figures 2(b), 3(c,d) of the Letter.

\end{document}